\bigskip
\documentclass[twocolumn,preprintnumbers,amsmath,amssymb]{revtex4}
\usepackage{graphicx}
\usepackage{fancyhdr}
\usepackage{dcolumn}
\usepackage{bm}
\usepackage{amssymb}
\usepackage{amsmath}
\usepackage{graphicx}
\usepackage{float}
\usepackage[normalem]{ulem}
\usepackage[dvips]{color}
\usepackage[colorlinks=true,citecolor=blue,linkcolor=black]{hyperref}

\begin{document}
\preprint{ }
\title{ Symmetry potentials and in-medium nucleon-nucleon
cross sections within the Nambu-Jona-Lasinio model in relativistic impulse
approximation}
\author{ Si-Na Wei\footnote{First author, 471272396@qq.com}, Rong-Yao Yang, Jing Ye,  Niu Li, Wei-Zhou Jiang\footnote{Corresponding author, wzjiang@seu.edu.cn}}
\affiliation{School of Physics, Southeast University, Nanjing
211189, China}
\begin{abstract}
In the relativistic impulse approximation (RIA),   we study
symmetry potentials and in-medium nucleon-nucleon (NN) cross sections with the Nambu-Jona-Lasinio (NJL) model that features chiral symmetry. The chiral symmetry that plays a fundamental role in the non-perturbative physics in the strong interaction  is
anticipated to add restrictive effects on the symmetry potentials
and in-medium NN cross sections. For comparison, we also perform
the study with the usual relativistic mean-field (RMF) model. The
numerical results with the NJL and RMF models are similar at
saturation density and below, since a priori fit was made to
saturation properties. With the increase of nuclear density, the
chiral symmetry starts to be restored partially in the NJL model,
resulting in the explicit fall of the scalar density. In a large
energy span, the symmetry potential acquires a significant rise
for the partial restoration of the chiral symmetry, compared to
the one with the RMF model. It is found that the in-medium NN
cross sections in the RIA with the NJL and RMF models both
increase with the density in the energy region interested in this
study, whereas those with the NJL model increase sharply as long
as a clear chiral symmetry restoration takes place. The different
tendency of observables in density can be transmitted to the
different energy dependence in the RIA. The NJL model is shown to
have a characteristic energy-dependent symmetry potentials and NN
cross sections beyond saturation point, apart from the RMF models.

\end{abstract}
\maketitle

\section{INTRODUCTION}

Heavy-ion reaction experiments are indispensable to extract the
nuclear equation of state (EOS) of asymmetric matter in various
density regions. In-medium nucleon-nucleon (NN) cross sections are
key inputs of reaction models, such as Glauber models~\cite{gl1}
and various transport models~\cite{gl2}. Up to now, though great
success in constraining the EOS of asymmetric matter has been
achieved~\cite{da02,li18}, considerable uncertainty of the EOS at
high densities remains due to the facts that the extraction of the
EOS from data relies actually on various models and that  the
solution of highly nonlinear nuclear in-medium problems is subject
to the many-body theory with various approximations. In
particular, the large uncertainty exists eminently in the density
dependence of the symmetry energy, which is one of two important
ingredients of nuclear EOS of asymmetric matter. In fact, the
uncertainty of the symmetry energy is closely associated with the
rather flexible isovector part of nucleon potentials, i.e., the
symmetry potential. Since the density dependence of the symmetry
energy plays a very important role in nuclear physics and
astrophysics~\cite{pa1,pa2,pa3}, the constraint on the uncertainty
in the symmetry energy and symmetry potential has received
enormous attention in past two
decades~\cite{li18,pa3,ts12,ho14,ba14,tr1,us2,us3,us1}. It has
been recognized that the theoretical uncertainty in extracting the
density dependence of the symmetry energy  from data of heavy-ion
reactions can be largely ascribed to the poor knowledge and
premature treatment of the isospin-dependent in-medium NN cross
sections~\cite{tr1}. For instance, in most Glauber calculations
only the isospin averaged free-space nucleon-nucleon cross section
is used. In deed, in spite of the well determined NN cross section
in free space,  the in-medium NN cross section is
phenomenological or  model dependent
~\cite{nn0,nn1,nn2,nn3,nn4,liu01,nn5,nn13,ji07,sa14,wa17,wa20}.

Due to the interplay between the uncertainty of the EOS and the
complexity of the in-medium interactions, it is far off
straightforward to derive simultaneously the symmetry potential
and in-medium NN cross sections microscopically or simply by model
extrapolations with consistency. For this reason, we seek help
of first principles with following two considerations. First,  we
will highly value the role of chiral symmetry in nuclear mean
fields and NN cross sections, since the chiral symmetry property
takes part to acquire most of the mass for nucleons and to
recover the chiral symmetry partially  in the dense medium.
Second, it is necessary to derive in-medium NN cross sections
consistent with the  mean fields of nuclear matter that determine
the symmetry potential. With these considerations, in this work we will deal with the symmetry potential and  in-medium NN cross
sections in the relativistic impulse approximation (RIA) that is
connected to the mean fields obtained from the chiral model.

In popular chiral models such as the linear-$\sigma$ and
Nambu-Jona-Lasinio (NJL) model~\cite{ch1,ch2,ch3}, the chiral
condensate on the nucleonic level is  related to the  scalar
field. Therefore, the partial restoration of chiral symmetry in
the nuclear medium brings the effect on  the scalar mean field and
consequently the in-medium NN cross sections. This will eventually
modify the symmetry energy extracted from heavy-ion reaction data.
In this work, we will adopt the NJL model to produce the mean
fields.  The original NJL model was established as a theory for
compound particles with constituent nucleons~\cite{ch2}. In the
past, the NJL model has been widely used on quark
level~\cite{nq1,nq2,nq3,nqj,nq4,nq13,nq14,nq15} and a nucleonic
level~\cite{nq5,nq6,nq8,nq9,nq10,nq0,nq11,nq12} as well.   On the
nucleonic level,  the original NJL model can't produce saturation
properties of nuclear matter. Similar to  the $\sigma$
model~\cite{ch3}, a scalar-vector interaction was introduced  to
cure this problem, and the NJL model has been extended to
include other interactions respecting the chiral
symmetry~\cite{nq10,nq0,nq11,nq12}.  We will start with the
nucleonic NJL model that can reproduce the saturation properties.

The RIA combines the Dirac decomposition of scattering amplitudes
and nuclear scalar and vector densities with the experimental
data. The optical potentials obtained from the RIA reproduced the
analyzing power and spin-rotation parameter in proton-nuclei
scatterings successfully~\cite{m1,m2,l1}, whereas the standard
nonrelativistic optical models failed to describe experimental
data for spin observables~\cite{a11,a22}. Later, the RIA was also
extended to the relative low-energy nucleon-nuclei elastic
scatterings~\cite{h1,h2,h4}. The RIA success interprets the
importance of the relativistic dynamics. In the past, the RIA  was
also used to study the symmetry
potentials~\cite{chen,li06,us2,us3} and in-medium NN cross
sections~\cite{ji07}. Although the scattering amplitudes can
provide a constraint on the symmetry potential in the RIA, it is
generally known that the the energy and density dependence of the
symmetry potential obtained from many other approaches relies
actually on specific parametrizations~\cite{pa3,us2,us3,us1}. As
the symmetry potential and in-medium NN cross sections diversify
for various in-medium interactions in nuclear models and
approaches, it is rather advantageous to study them with the RIA
in a unified manner.   In this work, we are thus motivated to
revisit the symmetry potential and in-medium NN cross sections
uniformly in the RIA based on the NJL model that follows the
chiral symmetry. For comparison, we will also perform the study
with the usual relativistic mean-field (RMF) model.

The paper is organized as follows. In Sec. \ref{RMF}, we briefly
introduce formalism and approaches used in this work. They include
a brief RMF formalism, the NJL model,  the RIA, and the in-medium
NN cross sections and relevant quantities.  Results and
discussions are  presented in Sec. \ref{results}. A summary is
finally given in Sec. \ref{summary}.

\section{Formalism}\label{RMF}
\subsection{Relativistic mean-field theory}
The original version of RMF model is the Walecka model\cite{wl1}
which only contains scalar and vector mesons  to provide the
medium-range attraction and short-range repulsion between the
nucleons, respectively. Since the  incompressibility of Walecka
model is too large, Boguta and Bodmer introduced the nonlinear
self-interactions of the $\sigma$  meson to bring down the
incompressibility reasonably\cite{wl2}. For studying the
isospin-asymmetric nuclei, the isovector ($\rho$) meson  was later
invoked to account  for the isospin dependence of the nuclear
force\cite{wl3}. To eliminate the scalar potential instability
caused by the nonlinear self-interaction of the $\sigma$ meson at high density, Bodmer and later Sugahara and Toki took the nonlinear
self-interaction of the $\omega$ meson into account\cite{wl4,st94}.
Due to the uncertainty of the symmetry energy,  the interaction
between the $\omega$   and $\rho$ meson was later involved in a
full Lagrangian~\cite{hp01}:
\begin{eqnarray}
\mathcal{L}&=&\bar{\psi}[\gamma_{\mu}(i\partial^{\mu}-g_{\omega}{\omega}^{\mu}-g_{\rho}\tau_3b^{\mu})-(M-g_{\sigma}\sigma)]\psi\nonumber\\
&&+
\frac{1}{2}(\partial_{\mu}\sigma\partial^{\mu}\sigma-m_{\sigma}^2\sigma^2)-\frac{1}{4}F_{\mu\nu}F^{\mu\nu}+\frac{1}{2}m_{\omega}^2{\omega}_{\mu}{\omega}^{\mu}
\nonumber\\
&&-\frac{1}{4}B_{\mu\nu}B^{\mu\nu}+\frac{1}{2}m_{\rho}^2b_{\mu}b^{\mu}-\frac{1}{3}g_2\sigma^3-\frac{1}{4}g_3\sigma^4\nonumber\\
&&
+\frac{1}{4}c_3({\omega}_{\mu}{\omega}^{\mu})^2+4\Lambda_Vg_{\rho}^2g_{\omega}^2{\omega}_{\mu}{\omega}^{\mu}b_{\mu}b^{\mu},
\label{nl}
\end{eqnarray}
where $g_{i}$ and $m_i$ with $ i=\sigma,\omega,\rho$  are the meson-nucleon coupling
constants and the meson masses, respectively.  $g_2$ and $g_3$ are the
coupling constants of the nonlinear self-interaction of the $\sigma$
meson. $c_3$ is the  coupling constant of the nonlinear
self-interaction of the $\omega$ meson.  $F_{\mu\nu}$ and $B_{\mu\nu}$ are the
strength tensors of the $\omega$ and $\rho$ mesons, respectively
\begin{eqnarray}
F_{\mu\nu}=\partial_{\mu}\omega_{\nu}-\partial_{\nu}\omega_{\mu},
B_{\mu\nu}=\partial_{\mu}b_{\nu}-\partial_{\nu}b_{\mu}.
\end{eqnarray}
With this Lagrangian, a nicely-fit RMF parametrization is  the
FSUGold by Todd and  Piekarewicz~\cite{wl5}. The equations of
motion for the nucleon and meson are actually the Euler-Lagrange
equations from above Lagrangian.    In the RMF approximation, they
are given as follow:
\begin{equation}
[i\gamma^{\mu}\partial_{\mu}-g_{\omega}\gamma^{0}
\omega_{0}-g_{\rho}\gamma^{0}b_{0}\tau_3-(M-g_{\sigma}\sigma)]\psi=0,
\label{qen1}
\end{equation}
\begin{eqnarray}
m_{\sigma}^2\sigma&=&g_{\sigma}\rho_S-g_2\sigma^2-g_3\sigma^3,
\label{qen2}\\
m_{\omega}^2 \omega_{0}&=& g_{\omega}
\rho_B-c_3\omega_{0}^3-8\Lambda_Vg_{\rho}^2g_{\omega}^2b_{0}^2{\omega}_{0}
,
\label{qen3}\\
 m_{\rho}^2 b_0&=&g_{\rho}
\rho_3-8\Lambda_Vg_{\rho}^2g_{\omega}^2{\omega}_{0}^2b_{0},
\label{qen4}
\end{eqnarray}
where $\rho_S=<\bar{\psi}\psi>$ and $\rho_B=<\bar{\psi}\gamma^0\psi>$
are the scalar  density and baryon density respectively.
$\rho_3=<\bar{\psi}\gamma^0\tau_3\psi>=\rho_p-\rho_n$ is the
difference between the proton and neutron densities. For a given
baryon density $\rho_B$ and isospin asymmetry parameter
$\delta=(\rho_n-\rho_p)/\rho_B$, the set of coupled equations can
be solved self-consistently by iterative methods.

\subsection{Nambu-Jona-Lasinio model of nucleonic level}
On the nucleonic level, the extended NJL which contains scalar,
vector and scalar-vector interactions can successfully produce the
saturation property~\cite{nq5}. For asymmetric nuclear matter, it
is reasonable to include the isovector and isovector-scalar
interactions\cite{nq10,nq0}. The Lagrangian is written as
\begin{eqnarray} \label{NJLQE1}
&&\mathcal{L}=\bar{\psi}(i\gamma_{\mu}{\partial^{\mu}}-m_0)\psi+\frac{G_S}{2}[(\bar{\psi}\psi)^2-(\bar{\psi}\gamma_5\tau\psi)^2]
\nonumber\\&&-\frac{G_V}{2}[(\bar{\psi}\gamma_{\mu}\psi)^2+(\bar{\psi}\gamma_{\mu}\gamma_5\psi)^2]\nonumber\\&&+
\frac{G_{\rho}}{2}[(\bar{\psi}\gamma_{\mu}\tau\psi)^2+(\bar{\psi}\gamma_{\mu}\gamma_{5}\tau\psi)^2]\\&&
+\frac{G_{SV}}{2}[(\bar{\psi}\psi)^2-(\bar{\psi}\gamma_5\tau\psi)^2][(\bar{\psi}\gamma_{\mu}\psi)^2+
(\bar{\psi}\gamma_{\mu}\gamma_5\psi)^2]\nonumber\\
&&+\frac{G_{\rho{S}}}{2}[(\bar{\psi}\gamma_{\mu}\tau\psi)^2+(\bar{\psi}\gamma_{\mu}\gamma_{5}\tau\psi)^2]
[(\bar{\psi}\psi)^2- (\bar{\psi}\gamma_5\tau\psi)^2]\nonumber,
\end{eqnarray}
where the $m_0$ is the bare nucleon mass, and $G_S$, $G_V$, $G_{SV}$,
$G_{\rho}$ and $G_{\rho S}$ are  the scalar, vector,
scalar-vector, isovector and isovector-scalar coupling constants,
respectively. In the RMF approximation, the Lagrangian can be
simplified to be
\begin{eqnarray} \label{eq5}
 \mathcal{L}&=&\mathcal{L}_{0}+\mathcal{L}_{IV}\nonumber\\
 &=&\bar{\psi}[i\gamma_{\mu}{\partial^{\mu}}-M^*- \gamma^0\Sigma]\psi-U,
\end{eqnarray}
where $M^*$, $\Sigma$, and $U$ are respectively defined as
\begin{eqnarray}
 &&M^*=m_0-(G_S+G_{SV}\rho_B^2+G_{{\rho}S}\rho_3^2)\chi_S, \label{eqgap} \\
 &&U=\frac{1}{2}(G_S\chi_{S}^{2}-G_V\rho_B^2-G_\rho\rho_{3}^2\nonumber\\&&+3G_{SV}\chi_{S}^{2}\rho_B^2
+3G_{{\rho}S}\rho_{3}^2\chi_S^2),\label{eq8}\\
&&\Sigma=G_V\rho_B+G_{\rho}\rho_3\tau_3-G_{SV}\chi_{S}^{2}\rho_B\nonumber\\
 &&-G_{{\rho}S}\rho_{3}\chi_S^2\tau_3,\label{eq7}
\end{eqnarray}
with
\begin{equation}\label{chi}
  \chi_S=-\sum_{i=p,n}\int_{k_{F_i}}^{\Lambda}\frac{d^3k}{(2\pi)^3} \frac{M^*}{\sqrt{k^2+{M^*}^2}}.
\end{equation}
Here, Eq.~(\ref{eqgap}) is the gap equation for the nucleon
effective mass   in  the NJL model, and the bare nucleon mass $m_0$ is obtained from the relation: $m_{\pi}^2f_{\pi}^2=m_0\chi_S^{vac}$.
In addition to the small $m_0$ that breaks the chiral symmetry explicitly, the dynamic mass, generated from the vacuum scalar condensate $\chi_S$,  breaks the chiral symmetry predominately, which is known as the spontaneous breaking of
chiral symmetry.  As the baryon density increases,
the Fermi momentum  will be approaching the momentum cutoff $\Lambda$, which is necessarily introduced to regularize the high-momentum behavior in the calculation, and the dynamic mass decreases  to  vanishing at the critical density with $k_F=\Lambda$ in symmetric matter where the chiral symmetry restoration takes place.
More details  of the chiral symmetry restoration in nuclear matter can be referred to Ref.\cite{nq8}.   In order to check the effect of  the chiral
restoration whose critical density is determined by the momentum
cutoff, we employ two different momentum cutoffs 350   and 400
MeV which correspond to the critical densities  2.37$\rho_0$ and
3.53$\rho_0$ ($\rho_0=0.16{\rm fm}^{-3}$), respectively. The
parameters for the NJL350 and NJL400 can be found in our previous
works~\cite{nq0,we18}. The incompressibility values are 262   and 296 MeV for NJL350 and NJL400, respectively. The symmetry energy and
the slope of symmetry energy with various momentum cutoffs are set
to be 31.6   and 50.0 MeV by adjusting $G_{\rho}$ and $G_{\rho
S}$, respectively.

\subsection{Relativistic impulse approximation}
In  proton-nucleus scattering, the scattering process can be
approximately treated   as incident proton scattered by each of
the nucleons in the target nucleus by neglecting the impact of
incident particle to the mean fields. Therefore, the Dirac optical
potential can be approximated as\cite{m1,m2}:
\begin{eqnarray}
\tilde{U}_{\rm opt}(q)=-\frac{4\pi ip_{\rm
lab}}{M}<\bar{\Psi}|\sum_{n=1}^{A}e^{i\vec{q}\cdot\vec{r}(n)}\hat{F}(q,n)|\Psi>,
\end{eqnarray}
where  $p_{\rm lab}$  and  $M$ are  the laboratory momentum and
mass of the incident  nucleon, respectively. $\Psi$ is the ground
state of target nucleus. $\hat{F}$ is Lorentz invariant NN
scattering amplitudes and can be  decomposed into five components:
\begin{eqnarray}\label{eqdecm}
\hat{F}&=&F_S+F_V\gamma_1^{\mu}\gamma_{2\mu}+F_T\sigma_1^{\mu\nu}\sigma_{2\mu\nu}\nonumber\\&&
+F_P\gamma_1^5\gamma_2^5+F_A\gamma_1^5\gamma_1^{\mu}\gamma_2^5\gamma_{2\mu},
\end{eqnarray}
where $F_S$, $F_V$, $F_T$, $F_P$ and $F_A$ are the scalar, vector,
tensor, pseudoscalar,  and axial vector amplitudes, respectively.
For the spin-saturated nucleus, $\hat{F}$ only contains scalar
($F_S$), vector ($F_V$), and tensor ($F_T$) terms. Since the
tensor term of  the scattering amplitude is small, the Dirac
optical potential can be approximately written as:
\begin{eqnarray}\label{eqria1}
\tilde{U}_{\rm opt}(q)=-\frac{4\pi ip_{\rm lab}}{M}
[F_S(q)\tilde{\rho}_S(q)+\gamma^0F_V(q)\tilde{\rho}_B(q)].
\end{eqnarray}
For a nucleon scattering off finite nuclei, the momentum transfer
$q$ between nucleon   and  finite nuclei is important for
understanding the scattering angle dependent physical
quantities(such as differential cross sections). When the target
is infinite nuclear matter,  the scalar and vector densities  are
constant in coordinate space. This means only the forward NN
scattering amplitudes ($F_S(q=0)$ and $F_V(q=0)$) survive.
Therefore the Dirac optical potential  is simplified
as~\cite{m1,m2}:
 \begin{eqnarray}\label{eqria2}
{U}_{\rm opt}=-\frac{4\pi ip_{\rm lab}}{M}
[F_S\rho_{S}+\gamma^0F_V\rho_B] \label{uop}.
\end{eqnarray}
 The forward NN elastic scattering amplitudes $F_S$ and $F_V$~\cite{m1,m2} have been determined directly  from the experimental NN phase
shifts~\cite{Ar83}, and have been successfully used to
describe pA elastic scattering with incident energies above 400
MeV~\cite{l1}. In this work, we apply these free-space NN elastic scattering amplitudes to perform the RIA study in the medium. As pointed out in Ref.~\cite{Ar83}, some resonance-like noise may exist in the data at nucleonic kinetic energies between 650 and 800 MeV. The RIA  can also project the inelastic noise approximately onto the Dirac optical potentials, since the inelastic scattering amplitudes, albeit involving a resonant state  mass larger than the nucleon counterpart, do not violate the Lorentz decomposition in Eq.(\ref{eqdecm}) and the RIA structure in Eqs.(\ref{eqria1}) and (\ref{eqria2}).  Consequently, our RIA results may include some inelastic ingredients from the data with the resonance-like noise in the certain energy region.
Below 400 MeV, one can use  the NN elastic scattering
amplitude of relativistic Love-Franey model developed by Murdock
and Horowitz~\cite{h1,h2}. $\rho_S$ and $\rho_B$ are the spatial
scalar and vector densities of infinite nuclear matter,
 \begin{eqnarray}
&&\rho_{S,i}=\int_0^{k_{Fi}}\frac{d^3k}{(2\pi)^3}\frac{M^*}{\sqrt{M^{*2}+k^2}},\nonumber\\
&&\rho_{B,i}=\frac{k_{Fi}^3}{3\pi^2},\mbox{ } i=n,p. \label{deo}
\end{eqnarray}
When the density dependent effective mass $M^*$ of nucleons is
obtained from nuclear models, the scalar density can be calculated
from Eq.(\ref{deo}) directly. The Dirac optical potential can be
expressed in terms of  scalar and vector optical potentials:
 \begin{eqnarray}
{U}_{\rm opt}&=&U_S^{\rm tot}+\gamma_0 U_0^{\rm tot},\nonumber\\
U_S^{\rm tot}&=&U_S+iW_S, \mbox{ } U_0^{\rm tot}=U_0+iW_0,
\label{uop2}
\end{eqnarray}
where $U_S$, $W_S$, $U_0$ and $W_0$ are respectively real scalar,
imaginary scalar, real vector and imaginary vector optical
potentials.

\subsection{Symmetry potential and in-medium NN cross section}
In order to study nuclear symmetry potential,  the nucleon mean
free path (MFP) and in-medium NN cross section, it is useful to
derive the Schr\"odinger equivalent potential (SEP) from the Dirac
equation. The Dirac equation can be decomposed into two equations
for the small and large components of the Dirac spinor. By
eliminating the small component of the Dirac spinor, one can
obtain a Schr\"{o}dinger-like equation for the large component. To
obtain the SEP, there are two schemes by  Jaminon et al~\cite{mj2}
and Hama et al~\cite{hama}. Regardless of the small Darwin term
and Coulomb potential, the SEP following the Jaminon's definition
(denoted by "J") is  written as~\cite{mj2}:
 \begin{eqnarray}
U_{\rm sep,J}^{\rm tot}=U_S^{\rm tot}+ U_0^{\rm tot}+
\frac{{U_S^{\rm tot}}^2- {U_0^{\rm tot}}^2}{2M}+\frac{U_0^{\rm
tot}E_{\rm kin}}{M}, \label{usep1}
\end{eqnarray}
with $U_{\rm sep,J}^{\rm tot}=U_{\rm sep,J}+iW_{\rm sep,J}$.
$E_{\rm kin}$ is the nucleon kinetic energy.   In the Hama's
definition~\cite{hama}, the SEP (denoted by "H") is given as:
 \begin{eqnarray}
U_{\rm sep,H}^{\rm tot}&=&[U_S^{\rm tot}+ U_0^{\rm tot}+
\frac{{U_S^{\rm tot}}^2- {U_0^{\rm tot}}^2}{2M}+\frac{U_0^{\rm
tot}E_{\rm kin}}{M}]\frac{M}{E}\nonumber\\&
=&\frac{1}{2E}[2EU_0^{\rm tot}+2MU_S^{\rm tot}+{U_S^{\rm tot}}^2- {U_0^{\rm tot}}^2], \label{usep2}
\end{eqnarray}
with $E=E_{kin}+M$.  Numerically, $U_{\rm sep,H}^{\rm tot}$ equals
$U_{\rm sep,J}^{\rm tot}M/E$. Eqs.(\ref{usep1}) and
(\ref{usep2})  can overestimate and underestimate the SEP at high
energies, respectively. As seen below, these two expressions
actually give rise to the the same nucleon MFPs and NN cross
sections.  By applying the SEP, the real and imaginary symmetry
potentials are obtained as:
 \begin{eqnarray}
&&U_{\rm sym}=\frac{U_{\rm sep}^{n}- U_{\rm sep}^{p}}{2\delta},
\nonumber\\&&  W_{\rm sym}=\frac{W_{\rm sep}^{n}-W_{\rm
sep}^{p}}{2\delta}, \label{usy}
\end{eqnarray}
where $U_{\rm sym}$ is the well-known Lane potential~\cite{mj3}.
Using the dispersion relation~\cite{mj4} and  assuming a complex
momentum $k = k_R + ik_I$,   the nucleon MFP can be
obtained as~\cite{ji07}:
\begin{eqnarray}
&&\lambda_{i}=\frac{1}{2k_I}=\frac{\sqrt{2}}{2}\{-(E^2-M^2-2MU_{\rm
sep}^i)+\nonumber\\&& [(E^2-M^2-2MU_{\rm sep}^i)^2+({2MW_{\rm
sep}^i})^2]^{1/2}\}^{-1/2}.\nonumber\\&& \label{mff}
\end{eqnarray}
Note that  $\lambda_{i,J}=\lambda_{i,H}$ because of the relation:
$U_{\rm sep,H}^{\rm tot}=U_{\rm sep,J}^{\rm tot}M/E$. Thus, we
neglect the subscript $J$ and $H$ in the following.   The nucleon
MFP can also be expressed  as the length of the unit volume
defined by the matter density and the NN cross section\cite{mj5}.
In this case, the nucleon MFP reads
 \begin{eqnarray}
\lambda_i=(\rho_p\sigma_{ip}^*+\rho_n\sigma_{in}^*)^{-1},
\label{lm1}
\end{eqnarray}
where $\sigma^*$'s are in-medium NN scattering cross sections,
$\rho_p$ and $\rho_n$ are the proton and neutron densities,
respectively. At some circumstances where the isospin dependence
is small, the in-medium NN cross sections can be obtained from the
above equation. It is useful to define two quantities~\cite{ji07},
\begin{eqnarray}
\tilde{\lambda}^{-1}=\frac{1}{2\delta}(\frac{1}{\lambda_n}-\frac{1}{\lambda_p}),
\mbox{ }
\tilde{\Lambda}^{-1}=\frac{1}{2}(\frac{1}{\lambda_n}+\frac{1}{\lambda_p}).
\label{lm2}
\end{eqnarray}
With Eqs.(\ref{lm1}) and  (\ref{lm2}), in-medium NN cross
sections can be obtained as:
\begin{eqnarray}
\sigma_{pp}^*=(\tilde{\Lambda}^{-1}+\tilde{\lambda}^{-1})/\rho_B,
\mbox{ }
\sigma_{np}^*=(\tilde{\Lambda}^{-1}-\tilde{\lambda}^{-1})/\rho_B,
\label{lm3}
\end{eqnarray}
by assuming $\sigma_{pp}^*=\sigma_{nn}^*$ on the fact that
$\tilde{\Lambda}$ and $\tilde{\lambda}$ were proved to be
approximately  independent of the isospin asymmetry of the medium
at high energies which is much larger than the Fermi
momentum~\cite{ji07}. The in-medium neutron-neutron and
proton-proton scattering cross sections are thus equal by
neglecting the small charge symmetry breaking effect~\cite{mj6}
and  small isospin dependent Pauli blocking effects  at high
energies in  asymmetric nuclear matter. Such an inference will
become weak at low energies or at high densities where the
isospin-dependent Pauli blocking arises  in $\tilde{\Lambda}$ and
$\tilde{\lambda}$ for the comparable Fermi momentum and kinetic
energy $E_{\rm kin}$.

\section{Results and discussions}
\label{results}

\begin{figure}[!htb]
\centering
\includegraphics[height=6cm,width=7cm]{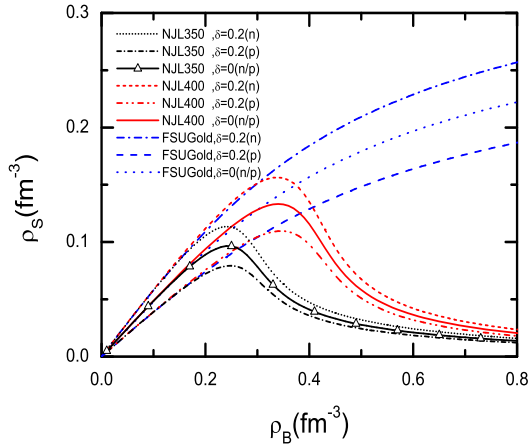}
\caption{The neutron and proton scalar densities as functions of
density with parametrizations of the NJL model and FSUGold. n and
p are abbreviations of neutron and proton, respectively, and  n/p
means neutron or proton at $\delta=0$.}\label{rhos}
\end{figure}

In RIA, the scalar and vector densities,  associated respectively
with the scalar and vector potentials  which represent the
relativistic dynamics, are two important components to work out
the Dirac optical potentials. The scalar density of nuclear matter
is calculated with Eq.(\ref{deo}) where  the nucleon effective
mass $M^*$ is obtained by solving the mean fields
self-consistently in relativistic models. Here, we adopt the NJL
model~\cite{nq0} and  a nicely-fit RMF parametrization
FSUGold~\cite{wl5} for comparison to check the effect of chiral
symmetry on observables. Fig.~\ref{rhos} shows the neutron and
proton scalar densities as functions of baryon density $\rho_B$.
The isospin asymmetry $\delta=(\rho_n-\rho_p)/\rho_B$ is taken to
be 0 and 0.2. At $\delta=0.2$, the neutron scalar density is
always above the proton scalar density for the higher Fermi
surface, as shown in Fig.~\ref{rhos}. We see that the scalar
density with the FSUGold increases with increasing the baryon
density. It is, however, interesting to find that the scalar
density with the NJL model first increases and then decreases with
increasing the baryon density. This result, distinct from common
nuclear models, is due to the characteristics of chiral symmetry
in the NJL model. With the increase of density, the chiral
symmetry undergoes gradual restoration which is reflected by a
more rapid dropping of the nucleon effective mass. When the
density is close to the critical density for chiral symmetry
restoration, the nucleon loses its mass down to the chiral limit,
and the scalar density of nuclear matter becomes small. Since the
critical density with the NJL350 is less than that with the
NJL400,  the scalar density with the NJL350 starts to decrease at
a density lower than the one with the NJL400.

\begin{figure}[!htb]
\centering
\includegraphics[height=7.5cm,width=7.5cm]{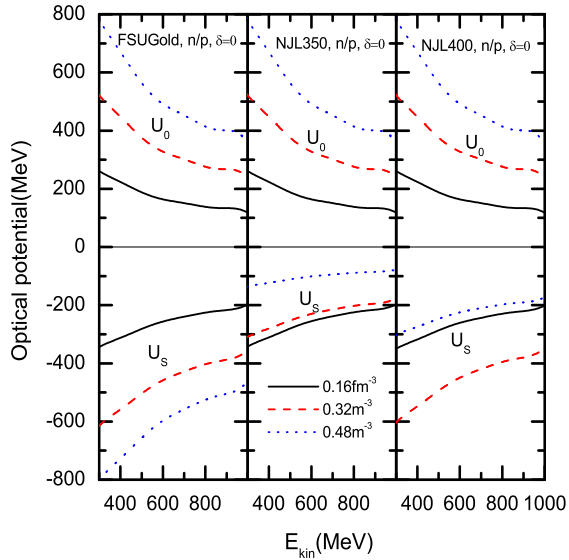}
\caption{Real  parts of the scalar ($U_S$) and vector ($U_0$)
Dirac optical potentials for neutrons and protons with various
parametrizations at $\rho_B=0.16,0.32,0.48 \mbox{ fm}^{-3}$ as
functions of the
 nucleon kinetic energy.}\label{opt1}
\end{figure}

With employing the above neutron and proton scalar densities and
the forward NN  scattering amplitudes, the scalar and vector Dirac
optical potentials for neutrons and protons in nuclear matter are
obtained directly by virtue of  Eq.(\ref{uop}) and
Eq.(\ref{uop2}).  In Fig.~\ref{opt1}, we illustrate the real parts
of the optical potentials at $\delta=0$ for densities
$\rho_B=0.16$,  $0.32$, and
$0.48\mbox{ fm}^{-3}$ with the NJL model and FSUGold. Note that
Neutron and proton Dirac optical potentials at $\delta\neq0$ are
reasonably separate for unequal proton and neutron scalar and
vector densities~\cite{chen}, while we do not depict them in
Fig.~\ref{opt1} for concision.  At $\rho_B=0.16\mbox{ fm}^{-3}$,
the Dirac optical potentials of various models (FSUGold, NJL350
and NJL400) are very close to each other, because present nuclear
models are well constrained by saturation properties. Beyond
saturation density, it is known that the large difference in the
nuclear EOS may arise from different models. In RMF models, the
difference in the EOS can partly from moderately different scalar
densities which is associated with the scalar potential. While at
high densities the dropping scalar densities given by the NJL
model are totally different from those given by usual RMF models
due to the chiral symmetry effect, the scalar component of the
optical potentials separates obviously from the one from the usual
RMF models. This can be observed clearly in Fig.~\ref{opt1}. At
$\rho_B=0.32\mbox{ fm}^{-3}$ which is close the critical density
in NJL350,    the scalar Dirac optical potential with the NJL350
shallows clearly, compared to that with the FSUGold and
NJL400. As the baryon density rises further to $0.48\mbox{
fm}^{-3}$, the chiral symmetry is already restored in NJL350, and
it is partially restored in NJL400. In this case, the NJL350 has
the very small scalar density, followed by the scalar density with
the NJL400. As a consequence, the scalar optical potential with
the NJL350 and NJL400 is clearly above the one with the FSUGold.
As shown in Fig.~\ref{opt1}, there is no difference in the vector
parts of Dirac optical potentials with the NJL350, NJL400 and
FSUGold, because the vector parts depend only on the same baryon
density. Note that the imaginary parts of the optical potentials
have the similar features as in the real parts, and we neglect the
figurations for simplicity.

\begin{figure}[!htb]
\centering
\includegraphics[height=11cm,width=8.5cm]{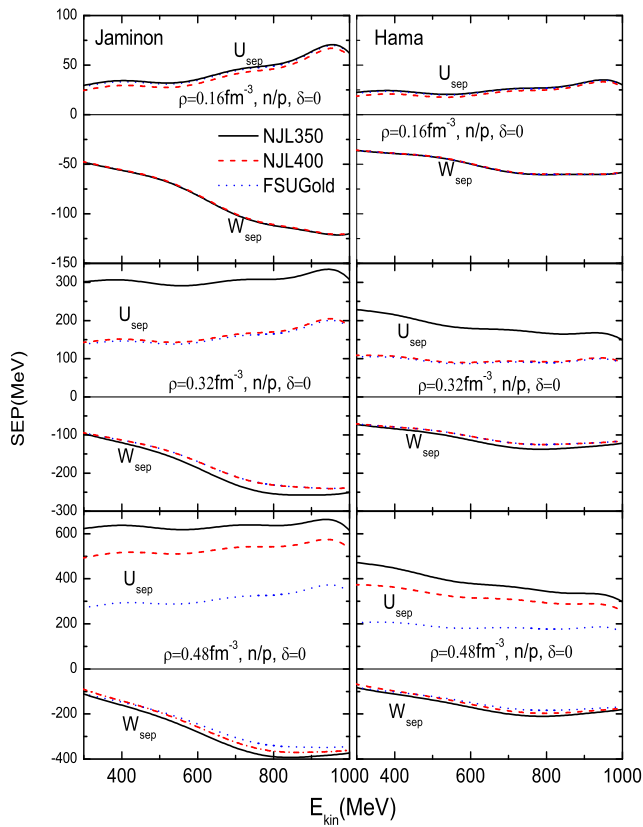}
\caption{Real and imaginary parts of Schr\"odinger equivalent
potentials as functions of nucleon kinetic energies in symmetric
matter at different densities. The real and imaginary parts are
above and below zero, respectively.  The results in the left and
right panels are calculated with Eq.(\ref{usep1}) and
Eq.(\ref{usep2}), respectively.}\label{usch}
\end{figure}

The nucleon SEPs, given  by Eq.(\ref{usep1}) and Eq.(\ref{usep2}),
are depicted in Fig.~\ref{usch} for symmetric nuclear matter with
$\delta=0$. We see that the nucleon SEPs are almost the same  at
$\rho_B=0.16\mbox{ fm}^{-3}$ with different models, which is
consistent with the case for optical potentials, as shown in
Fig.~\ref{opt1}. The effect from the chiral symmetry starts to
arise as the density increases. For instance, at
$\rho_B=0.32\mbox{ fm}^{-3}$, since the chiral symmetry in NJL350
is nearly restored, the real part of the nucleon SEP increases
sharply due to the large restoration of chiral symmetry. Similar
case takes place for the real part of the nucleon SEP with the
NJL400 at $\rho_B=0.48\mbox{ fm}^{-3}$ which is close to its
critical density of chiral symmetry restoration.  As shown in
Fig.~\ref{usch}, the imaginary part of the nucleon SEP with the
NJL model becomes lower than the one with the FSUGold, resulting
from  the effect of  the chiral symmetry restoration.  In general,
the density and energy evolution of $U_{\rm sep,H}^{\rm tot}$ and
$U_{\rm sep,J}^{\rm tot}$ is similar in two cases. However, since
a factor $M/E$ exists in the relation $U_{\rm sep,H}^{\rm
tot}=U_{\rm sep,J}^{\rm tot}M/E$, a shrinkage of  the SEPs in
right panels appears in comparison to those in left panels.
\begin{figure}[!htb]
\centering
\includegraphics[height=8cm,width=8cm]{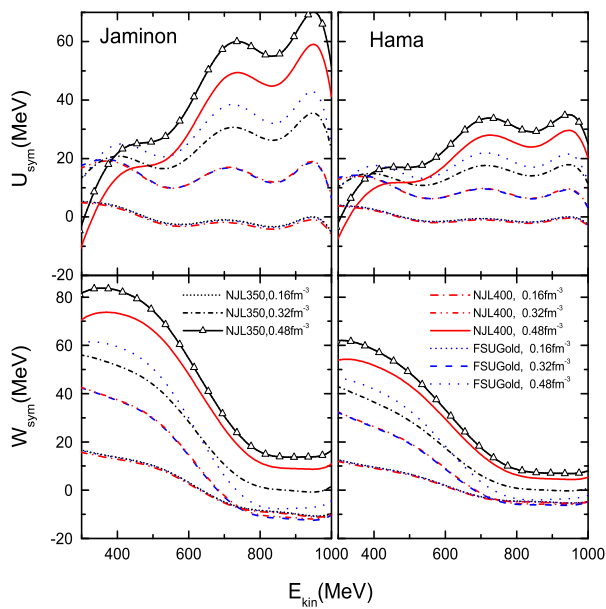}
\caption{The symmetry potential  as a function of the nucleon
kinetic energy at  different densities.  The differentiation of the
left and right panels is similar to that in Fig.~\ref{usch}.
}\label{usym}
\end{figure}
\begin{figure}[!htb]
\centering
\includegraphics[height=8cm,width=8cm]{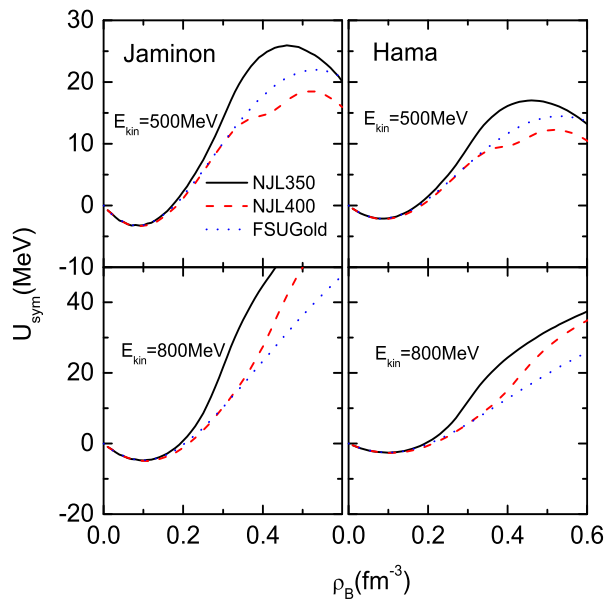}
\caption{The symmetry potential as a function of density at
$E_{\rm kin}=500$   and 800 MeV.   }\label{usym1}
\end{figure}

The symmetry potential can be simply regarded as the potential
part of the symmetry energy, and the uncertainty of the symmetry
energy is largely attributed to that of the symmetry
potential~\cite{chen,h2,li06,us1,h4}. However, the symmetry
potential in RIA is rather constrained by the scattering amplitude
together with densities from relativistic models. From different
sources of optical potentials, the symmetry potential is extracted
to be linear in the kinetic energy in the energy region below 200
MeV~\cite{us1}. With the RIA, the linear dependence on the nucleon
kinetic energy was approved earlier by Li et al~\cite{li06}. As
the kinetic energy  extends to the region above 400 MeV, the
symmetry energy stays almost unchanged for densities in the
vicinity of saturation~\cite{chen}.  In Fig.\ref{usym}, we show
the real and imaginary symmetry potential, given by
Eq.(\ref{usy}), as functions of nucleon kinetic energy. It is
found that the $U_{\rm sym}$ with the  NJL350, NJL400 and FSUGold
are close to each other at $\rho_B=0.16\mbox{ fm}^{-3}$, and it is
nearly constant at higher nucleon kinetic energy, consistent with
that in Ref.~\cite{chen}. With the increase of density, the chiral
symmetry effect on the symmetry potentials is just similar to that
for the SEP, as can be observed by comparing Fig.~\ref{usym} with
Fig.~\ref{usch}. The difference is that at high densities the real
symmetry potential, namely the Lane potential $U_{\rm sym}$,
fluctuates with energy that is seeded by the wavy $F_V(E_{\rm
kin})$ in energy and the amplification by multiplying $E_{\rm
kin}$ in the last term in Eq.(\ref{usep1}). We can infer that the
fluctuation is dominated by the vector potential, since it
survives in the NJL results whose scalar potential is negligible
near the critical density. This can be nicely  verified
numerically. It should be noted that the imaginary symmetry
potential displays an energy-dependence different from that of the real symmetry potential at a fixed density: It decreases with
increasing nucleon kinetic energy. Nevertheless, the symmetry
potential is largely modified by the effect of chiral symmetry
restoration especially at higher energies, as shown in
Fig.~\ref{usym}.  Yet, we see that  the symmetry potential relies
on the approximations adopted. The shallower symmetry potentials
correspond to  the shallower SEPs as shown in Fig.~\ref{usch}.

 Shown in Fig.~\ref{usym1} is the density profile
of the Lane potential for $E_{\rm kin}$=500 and 800 MeV. It is
clearly shown that the symmetry potential becomes apart from
different models with increasing the density. Interestingly, when
the chiral symmetry is partially restored at higher densities, the
separation between those from different parametrizations appears
to be large. Especially at $E_{\rm kin}=800$ MeV, the separation
actually characterizes the difference induced by the chiral
symmetry restoration in the NJL model. In deed, the difference
between the NJL model and the RMF FSUGold is irrelevant to the
vector potentials at a fixed density and a fixed energy, and it is
uniquely associated with the scalar potential (or, the scalar
density) and the nucleon effective mass therein. However, the
symmetry potential with the FSUGold   at $E_{\rm kin}=500$ MeV is
unusually higher than the one with the NJL400 at higher densities,
as shown in the upper panel of Fig.~\ref{usym1}. In fact, the
uplift of the symmetry potential  at lower energies with the
FSUGold is ascribed to the cancellation between the
energy-dependent terms in Eq.(\ref{usep1}): $U_S^{\rm tot}$ and
${U_S^{\rm tot}}^2/2M$ that are opposite in sign. Though the
values of $U_S^{\rm tot}$ and ${U_S^{\rm tot}}^2/2M$ are
individually much larger than those given by the NJL model at
higher densities, the cancellation can give rise to a value
comparable to that with the NJL model. Note that this unusual
uplift can  be found consistently in Fig.~\ref{usym} in the
low-energy region. With increasing the energy, the clearly lower
optical potentials do not favor a delicate cancellation anymore,
resulting in a large difference induced by the chiral symmetry
effect, as shown in Figs.~\ref{usym} and \ref{usym1}, even though
the  SEPs are suppressed in the Hama's expression with increase of
the kinetic energy, see Eq.(\ref{usep2}).
\begin{figure}[!htb]
\centering
\includegraphics[height=7.5cm,width=7.5cm]{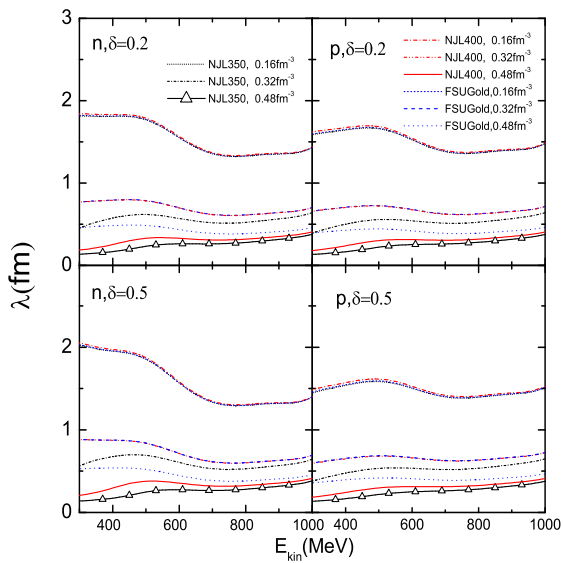}
\caption{The nucleon mean free path as a function of the nucleon
kinetic energy for different densities  with the NJL350, NJL400
and FSUGold.}\label{fmfp}
\end{figure}
\begin{figure}[!htb]
\centering
\includegraphics[height=9cm,width=7cm]{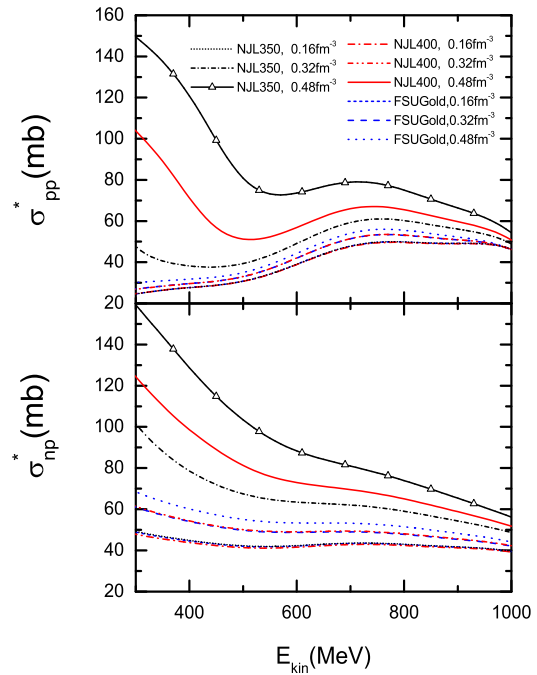}
\caption{ The in-medium NN   cross section   as a function of the nucleon kinetic energy at densities 0.16, 0.32 and 0.48
fm$^{-3}$.}\label{fcro}
\end{figure}

\begin{figure}[!htb]
\centering
\includegraphics[height=6.5cm,width=7.5cm]{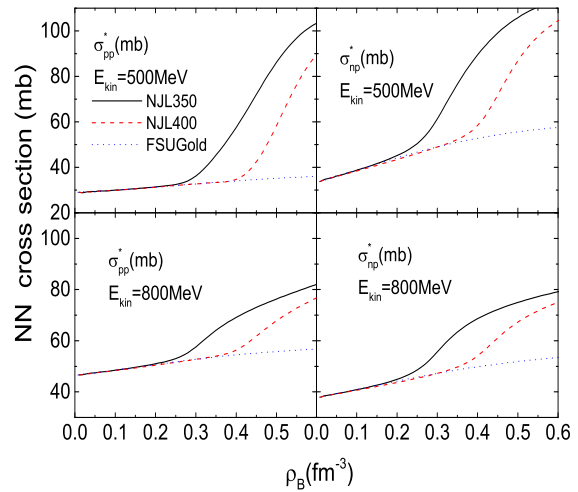}
\caption{ The in-medium NN  cross section   as a function of
density at $E_{\rm kin}=500$   and 800 MeV. }\label{dcro}
\end{figure}
Before addressing the in-medium NN cross sections, we present in
Fig.~\ref{fmfp} the nucleon MFPs at $\delta=0.2$ and 0.5,
calculated by Eq.(\ref{mff}).   The MFP of a particle is  the
average distance travelled between two successive collisions of
the particle, and here it is an input to work out the in-medium NN
cross section. The nucleon MFP at high densities will be obviously
shorter than that at low densities, as is consistently given in
Fig.~\ref{fmfp}. It is worthy to note that the restoration of
chiral symmetry leads to a decrease in MFPs, caused by a sharp
increase in the real SEP. For instance, at $\rho_B=0.32\mbox{
fm}^{-3}$ where the chiral symmetry in NJL350 is largely restored,
the nucleon MFPs with the NJL350 are lowered clearly, compared to
those with the NJL400 and FSUGold. As shown in Fig.~\ref{fmfp},
the high-energy behavior of the MFPs tends to be flattened in all
of present parametrizations, although the MFPs are not a
monotonous function of nucleon kinetic energy especially at high
densities in the NJL model. Besides, the nucleon MFPs are
dependent on the isospin, see Eq.(\ref{mff}). With increasing
$\delta$, the difference between neutron and proton MFPs becomes
moderately larger~\cite{ji07},  as shown in Fig.~\ref{fmfp}.

Now, we discuss the in-medium NN cross sections. Shown in
Fig.~\ref{fcro} is the in-medium NN   scattering cross sections as a
function of kinetic energy with NJL350, NJL400 and FSUGold. Once
again, we see that the overlap of the curves at saturation density
for the well constrained saturation properties of these models. As
shown in the figure, the NN cross sections with the FSUGold just
rise moderately with increasing the density, while the rise with
the NJL350 and NJL400 can be dramatic especially at large
densities close to the critical density. The large rise with the
NJL model can be understood by the small scalar densities which
yield small nucleon effective masses. Phenomenologically, the
in-medium NN cross sections were approximately scaled down by the
nucleon mass square in the medium~\cite{tr1} at low energies,
whereas here the NN cross sections evolve upwards with decreasing
the nucleon effective mass $M^*$. In the kinetic energy region
where the RIA applies, the lower energy region displays a more
apparent rise of $\sigma_{pp}^*$ and $\sigma_{np}^*$ with
increasing the density, while the contraction of the rise at
higher energies is consistent with the correspondingly reduced
scattering amplitudes. As seen in Eq.(\ref{lm3}), there is a
difference between the $\sigma_{np}^*$ and  $\sigma_{pp}^*$. It is
known that at low energies less than 300 MeV, the free
$\sigma_{np}$ is 3 times the free $\sigma_{pp}^*$. The proportion
between the neutron-proton and proton-proton cross sections in the
medium changes with the density and energy. As seen in
Fig.~\ref{fcro}, the in-medium cross section $\sigma_{pp}^*$
develops to be above $\sigma_{np}^*$ at higher energies.

The density dependence of the NN   cross sections are shown in
Fig.~\ref{dcro} for  $E_{\rm kin}=500$ MeV and 800 MeV. It is
striking that the sharp difference in the NN cross sections
appears for different models with increasing the density. The
difference in the NN cross sections between different models is
given by the scalar component of the optical potentials, since the
vector optical potential in RIA  is identical  for different
models at a fixed density and a fixed energy. As seen in the left
upper panel of Fig.~\ref{dcro}, the NN cross section with the NJL
model can be larger than that with the FSUGold by a factor 3 at
high densities. The results in Fig.~\ref{dcro} reflect  a fact
that the chiral  symmetry restoration has a direct and clear
correlation with  the in-medium NN cross sections: The eclipse of
the nucleon mass caused by the chiral symmetry restoration yields
an apparently large rise in the in-medium NN cross sections.
Not only can the chiral symmetry constrain the strong interaction,
it can also signal the internal structures of nucleons which might
be encountered in the energetic heavy-ion collisions. In deed, the
original NJL model and later versions treated the hadrons as
composite constituents. It has been recognized generally that the
chiral phase transition coincides with the color deconfinement,
evidenced by the vanishing baryon mass (from constituent mass to
the current mass)~\cite{CO80}, although quarkyonic matter between the hadron and quark phases
was proposed in the recent past~\cite{DF06,Mc07}. Similar
role of the chiral symmetry in pointing to the internal structure
ends but not in the more fundamental structure of quarks and
leptons by solving the axial anomaly~\cite{PA75,TH80}. Empirically
speaking, the effect produced by the chiral symmetry may roughly
reflect the influence of the interior quark degrees of freedom in
dense matter. In this sense, the consideration of chiral symmetry
in mean-field potentials is of significance and economic for the
transport models when reaching out to internal degrees of freedom
in the high density region. In deed, the high density region where
the chiral symmetry matters was produced by the energetic
heavy-ion collisions, for instance, see Ref.~\cite{da02,Re12}. We
notice that the present simple RIA should not be perfect when the
in-medium effect of the pion production resonances are treated
equally as the nucleons. In the low energy region below the  pion
production threshold, one can deal with this issue better by
extending the RIA to  the relativistic Love-Franey
model~\cite{h1,h2,h4,li06}.

We should figure out that the increasing trend of in-medium NN
cross sections with the density at high energies is different from
that in the literature at low nucleon
energies~\cite{nn0,nn1,nn2,nn3,liu01,nn13}  and that the
significance of this increasing trend is based more upon the
comparison to the results with the usual RMF model (FSUGold).   It
was familiar inadvertently that the in-medium NN cross sections
decrease with increasing the density at low energies. Our finding
in the high energy region, albeit different, is not unique. In
fact, it was found in the Dirac-Brueckner theory that the
in-medium NN cross sections can increase with the density at high
nucleon momenta~\cite{sa14}. Note that the present relativistic
treatment can result in the proximity of the NN cross sections at
very low densities to the free-space limits, consistent with that
in Ref.~\cite{ji07}.

At last, it is worthy to note that the symmetry potential effect
on  the in-medium NN cross sections is not included in the present
work. The symmetry potential just has a negligible effect around
the saturation density~\cite{ji07}. In much denser matter where
the RMF works nicely, the role of  the symmetry potential that is
associated with the isospin-dependent Pauli blocking would be
limited in the in-medium NN cross section in the case of $E_{\rm
kin}>k_F$. However, to better work out the NN cross sections in
asymmetric matter, we need to develop a method to include the
contribution of the symmetry potential and leave it in a future
work.

\section{Summary}
\label{summary} In this work, we have studied the effect of the
chiral symmetry effect in the NJL model on  the symmetry potential
and in-medium NN  cross section based on the Dirac optical
potentials in RIA. In comparison, we perform the calculation with
the RMF parametrization FSUGold.  We have calculated the Dirac
optical potentials, the Schr\"odinger equivalent potentials,
symmetry potentials, nucleon mean free paths, and the in-medium NN
cross sections with these models. We have found that the key
ingredient in RIA is the scalar density, as the vector component
is identical for all models at a fixed density and a fixed nucleon
kinetic energy. The scalar density obtained from the NJL model is
dramatically reduced at higher densities in comparison to that
with usual RMF models, because the chiral restoration in the
medium leads to a vanishing nucleon effective mass in the NJL
model. We have observed that the physical quantities mentioned
above undergo a consistently similar variation with the evolution
of nucleon density and kinetic energy.

With the chiral symmetry being restored considerably, the real
part of SEPs increases sharply, and this sharp rise is transmitted
to the symmetry potential.  For high kinetic energies, the chiral
symmetry effect on the symmetry potential can be clearly measured
by the difference  with the FSUGold.   The nucleon MFPs
with the chiral symmetry model are obviously lower than those with
the FSUGold, as the chiral symmetry starts to be restored
partially. Correspondingly, the in-medium NN cross sections  with
the NJL model increases sharply for the large eclipse of the
nucleon effective mass caused by the chiral symmetry restoration.
In short, the in-medium NN cross sections in RIA can measure the
chiral symmetry restoration. On the other hand, the in-medium NN
cross sections as a key input of the transport models should
account for the chiral symmetry effect in simulating the creation
of dense matter.

\section*{ACKNOWLEDGMENT}
The work was supported in part by the National Natural Science
Foundation of China under Grant No. 11775049.


\begin{thebibliography}{99}
\bibitem{gl1}I. Tanihata et al., a special volume of Nucl. Phys. A \textbf{693}, 1 (2001).
\bibitem{gl2} J. Xu, Prog. Part.  Nucl. Phys. \textbf{106}, 312
(2019).
\bibitem{da02} P. Danielewicz, R. Lacey, and W. G. Lynch,
Science \textbf{298}, 1592 (2002).
\bibitem{li18}B. A. Li, B. J. Cai, L. W. Chen, and J. Xu, Prog. Part.  Nucl. Phys.
\textbf{99}, 29 (2018).


  \bibitem{pa1} J. M. Lattimer, M. Prakash, Science \textbf{304},  536 (2004).
 \bibitem{pa2} A. W. Steiner, M. Prakash, J. M. Lattimer, P. J. Ellis, Phys. Rep. \textbf{411},  325 (2005).
\bibitem{pa3} B. A. Li, L. W. Chen, C. M. Ko, Phys. Rep.
\textbf{464}, 113 (2008).

 \bibitem{ts12}M. B. Tsang, J. R. Stone, F. Camera,  P. Danielewicz,  S. Gandolfi, K. Hebeler, et al., Phys. Rev. C \textbf{86}, 015803 (2012).
\bibitem{ho14}C. J. Horowitz, E. F. Brown, Y. Kim, et al., J.
Phys. G \textbf{41}, 093001 (2014).
 \bibitem{ba14}B. A. Li, A. Ramos, G. Verde, I. Vida\~na (Editors), Top-
ical issue on nuclear symmetry energy, in The European Physical
Journal A, Vol. \textbf{50} (2014).
 \bibitem{tr1} B. A. Li and L. W. Chen, Phys. Rev. C \textbf{72}, 064611 (2005).

\bibitem{us2}R. Chen, B. J. Cai, L. W. Chen, B. A. Li, X. H. Li, and C. Xu, Phys. Rev. C \textbf{85}, 024305 (2012).
\bibitem{us3}R. Wang, L.W. Chen, and Y. Zhou, Phys. Rev. C \textbf{98}, 054618 (2018).
\bibitem{us1}C. Xu, B. A. Li, and L. W. Chen, Phys. Rev. C \textbf{82}, 054607 (2010).

\bibitem{nn0}D. Klakow, G. Welke, and W. Bauer, Phys. Rev.
C \textbf{48}, 1982 (1993).
\bibitem{nn1} G. Q. Li and R. Machleidt, Phys. Rev. C \textbf{48}, 1702 (1993).
\bibitem{nn2} G. Q. Li and R. Machleidt, Phys. Rev. C \textbf{49}, 566 (1994).
\bibitem{nn3} P. Danielewicz, Nucl. Phys. A \textbf{673}, 375 (2000).

\bibitem{nn4} C. Fuchs, A. Faessler, and M. El-Shabshiry, Phys. Rev. C \textbf{64}, 024003 (2001).

\bibitem{liu01}J. Y. Liu, W. J. Guo, S. J. Wang, W. Zuo, Q. Zhao,
Y.F. Yang, Phys. Rev. Lett. \textbf{86}, 975 (2001).
\bibitem{nn5} F. Sammarruca and P. Krastev, Phys. Rev. C \textbf{73}, 014001 (2006).

\bibitem{nn13} H. F. Zhang, Z. H. Li, U. Lombardo, P. Y. Luo, F. Sammarruca, and W. Zuo, Phys. Rev.
C \textbf{76}, 054001 (2007).
\bibitem{ji07}W. Z. Jiang, B. A. Li, and L. W. Chen, Phys. Rev. C \textbf{76}, 044604 (2007).

\bibitem{sa14}F. Sammarruca, Eur. Phys. J. A \textbf{50}, 22 (2014).
\bibitem{wa17} T. T.  Wang,  Y. G.  Ma,  C. J.  Zhang,  Z. Q.  Zhang,  Phys.
Rev. C \textbf{97}, 034617 (2018).
\bibitem{wa20}R. Wang, Z. Zhang, L. W. Chen, C. M. Ko, Y. G. Ma, Phys. Lett. B \textbf{ 807},
135532 (2020).
\bibitem{ch1} M. Gell-Mann, M. L\'{e}vy, Nuovo Cimento \textbf{16}, 705 (1960).
\bibitem{ch2} Y. Nambu, G. Jona-Lasinio, Phys. Rev. \textbf{122}, 345 (1961).
\bibitem{ch3} J. Boguta, Phys. Lett. B \textbf{120},  34 (1983).


\bibitem{nq1}  S. P. Klevansky, Rev. Mod. Phys. \textbf{64}, 649 (1992) .
\bibitem{nq2}  M. Buballa, Phys. Rep. \textbf{407}, 205 (2005).
\bibitem{nq3} V.A. Miransky, M. Tanabashi, and K. Yamawaki, Phys. Lett.
B \textbf{221}, 177 (1989).
\bibitem{nqj} W. Z. Jiang, X. J. Qiu, Z. Y. Zhu, and Z. J. He,
 Phys. Rev. C \textbf{65}, 015210 (2001).
\bibitem{nq4} K. Fukushima, Phys. Lett. B \textbf{591}, 277 (2004).
 \bibitem{nq13} P. Costa, H. Hansen, M. C. Ruivo, and C. A. de Sousa, Phys.
Rev. D \textbf{81}, 016007 (2010).
\bibitem{nq14}H. Kohyama, D.
Kimura, and T. Inagaki, Nucl. Phys. B
 \textbf{896}, 682 (2015).
\bibitem{nq15} C. M. Li, J. L. Zhang, Y. Yan,Y. F. Huang, and H. S. Zong, Phys. Rev. D \textbf{97}, 103013
(2018).

\bibitem{nq5} V. Koch, T. S. Biro, J. Kunz, U. Mosel, Phys. Lett. B \textbf{185}, 1 (1987).
\bibitem{nq6}  U. Vogl, W. Weise, Prog. Part. Nucl. Phys. \textbf{27},  195 (1991).

\bibitem{nq8} I. N. Mishustin, L. M. Satarov, W. Greiner, Phys. Rep. \textbf{391},  363 (2004).
\bibitem{nq9} T. G. Lee, Y. Tsue, J. da Providencia, C. Providencia, M. Yamamura, Prog. Theor.
Exp. Phys. \textbf{2013}, 013D02 (2013); 1207.1499 [hep-ph].

\bibitem{nq10} H. Pais, D. P. Menezes, C. Providencia, Phys. Rev. C \textbf{93}, 065805 (2016).

\bibitem{nq0}S. N. Wei, W. Z. Jiang, R. Y. Yang, and D. R. Zhang, Phys. Lett. B \textbf{763}, 145 (2016).
\bibitem{nq11} C. A. Graeff, M. D. Alloy, K. D. Marquez, C. Providencia, and D. P. Menezes, J. Cosmol. Astropart. Phys. \textbf{01}, 024 (2019).
\bibitem{nq12} Y. J. Chen, Chin. Phys. C \textbf{43}, 035101 (2019).

\bibitem{m1} J. A. McNeil, L. Ray, and S. J. Wallace, Phys. Rev. C \textbf{27}, 2123 (1983).
\bibitem{m2} J. A. McNeil, J. R. Shepard, and S. J. Wallace, Phys. Rev. Lett. \textbf{50}, 1439 (1983).
\bibitem{l1} L. Ray and G. W. Hoffmann, Phys. Rev. C \textbf{31}, 538 (1985).

\bibitem{a11} J. P. Auger, J. Gillespie, and R. J. Lombard, Nucl. Phys. A \textbf{212}, 372 (1976).
\bibitem{a22}B. Klem, G. Igo, R. Talaga, et.al., Phys. Bev. Lett. \textbf{38}, 1272 (1977).

  \bibitem{h1} C. J. Horowitz, Phys. Rev. C \textbf{31}, 1340 (1985).
\bibitem{h2} D. P. Murdock and C. J. Horowitz, Phys. Rev. C \textbf{35}, 1442 (1987).

\bibitem{h4}Z. P. Li, G. C. Hillhouse, and J. Meng, Phys. Rev. C \textbf{77}, 014001 (2008).


\bibitem{li06}Z. H. Li, L.W. Chen, C. M. Ko, B. A. Li, and H. R. Ma, Phys. Rev. C \textbf{74}, 044613 (2006).

 \bibitem{chen}L. W. Chen, C. M. Ko, and B. A. Li, Phys. Rev. C \textbf{72}, 064606
(2005).



\bibitem{wl1} J. D. Walecka, Ann. Phys. \textbf{83},491 (1974).
\bibitem{wl2}  J. Boguta and A. R. Bodmer, Nucl. Phys. A \textbf{292}, 413 (1977).
  \bibitem{wl3} B. D. Serot, Phys. Lett. B \textbf{86}, 146 (1979).
  \bibitem{wl4} A. R. Bodmer, Nucl. Phys. A \textbf{526 }, 703(1991)
  \bibitem{st94}Y. Sugahara and H. Toki, Nucl. Phys. A \textbf{579}, 557 (1994).
  \bibitem{hp01}C. J. Horowitz and J. Piekarewicz, Phys. Rev. Lett. \textbf{86}, 5647 (2001).
   \bibitem{wl5} B. G. Todd-Rutel and J. Piekarewicz, Phys. Rev. Lett.  \textbf{95}, 122501  (2005).

\bibitem{we18}S. N. Wei, R. Y. Yang, and  W. Z. Jiang, Chin. Phys. C \textbf{42}, 054103 (2018).
\bibitem{Ar83}R. A. Arndt, L. D. Roper, R. A. Bryan, R. B. Clark, B. J. VerWest, and P. Signell, Phys. Rev. \textbf{D} 28, 97(1983).


\bibitem{mj2} M. Jaminon, C. Mahaux, and P. Rochus, Nucl. Phys. A \textbf{365}, 371
(1981).
\bibitem{hama}S. Hama, B. C. Clark,  E. D. Cooper, H. S. Sherif, and R. L. Mercer, Phys. Rev. C \textbf{41},  2737-2755 (1990).
\bibitem{mj3}A. M. Lane, Nucl. Phys. \textbf{35}, 676 (1962).
 \bibitem{mj4}G. Q. Li, R. Machleidt, and Y. Z. Zhuo, Phys. Rev. C \textbf{48}, 1062
(1993).
 \bibitem{mj5}V. R. Pandharipande, S. C. Pieper, Phys. Rev. C \textbf{45}, 791 (1992).

 \bibitem{mj6} G. Q. Li and R. Machleidt, Phys. Rev. C \textbf{58}, 1393 (1998); \textbf{58},
3153 (1998).
\bibitem{CO80}S. R. Coleman and E. Witten, Phys. Rev. Lett. \textbf{45}, 100 (1980).
\bibitem{DF06}P. de Forcrand and S. Kratochvila, Nucl. Phys. \textbf{B} (Proc. Suppl.) 153, 62
(2006).
\bibitem{Mc07}L. McLerran and R. D. Pisarski, Nucl. Phys. \textbf{A} 796,83 (2007).
\bibitem{PA75}J. C. Pati, A. Salam, and J. Strathdee,
Phys. Lett. \textbf{B} 59, 265(1975).
\bibitem{TH80} G. t'Hooft, in
Recent Development in Gauge Theories (Plenum Press, New York,
1980).
\bibitem{Re12}W. Reisdorf,Y.Leifels, A. Andronic, R. Averbeck V. Barret et al., Nucl. Phys. \textbf{A}876, 1 (2012).

\end{thebibliography}
\end{document}